\documentclass[a4paper,11pt]{article}
\usepackage{pos}
\usepackage{xspace}

\input{defs.tex}

\title{Correlations between \Upsn production and the underlying event in \pps collisions with the ATLAS detector}

\author[]{Baidyanath Sahoo$^{*}$ On behalf of the ATLAS Collaboration}

\affiliation[*]{Department of Particle Physics and Astrophysics, Weizmann Institute of Science,\\
Herzl Street 234, Rehovot 7610001, Israel}

\emailAdd{baidya.wis@gmail.com, baidyanath.sahoo@weizmann.ac.il}

\abstract{Recent studies from various experiments at the LHC and RHIC suggest that small collision systems exhibit quark-gluon plasma (QGP) like signatures that were typically observed in large systems. While QGP like signatures have been observed in the soft sector, strong constraints on possible partonic energy loss were established in the hard sector.

At the same time, there are multiple pieces of evidence pointing to correlations between hard and soft particle production in proton-proton collisions. The correlations between hard and soft particle production in $pp$ collisions are studied in events with $\Upsilon$ mesons, which provide a valuable hard probe to study such effects. The multiplicity and distributions of kinematic variables of charged particles produced in association with an $\Upsilon$ meson are measured using 139 fb$^{-1}$ of data at $\s = 13$~TeV collected by the ATLAS experiment at the LHC. The measurement is performed separately for the first three $\Upsilon$(nS) states in intervals of their transverse momentum. At low transverse momentum, the associated charged-particle multiplicity is observed to be systematically lower in collisions where $\Upsilon$(2S) or $\Upsilon$(3S) is produced compared to $\Upsilon$(1S), with the difference decreasing as the $\Upsilon$ transverse momentum increases.}

\FullConference{%
  33rd International Workshop on Deep Inelastic Scattering and Related Subjects (DIS2026)\\
  4-8 May 2026\\
  Bologna, Italy
}

\begin{document}

\providecommand{\logo}{}
\renewcommand{\logo}{}
\providecommand{\posurl}{}
\renewcommand{\posurl}{}

\maketitle

\section*{Introduction}
Correlations between hard and soft particle production in proton-proton (\pps) collisions provide information on the interplay between hard scattering and soft QCD dynamics. The underlying event (UE) comprises particle activity accompanying the hard scattering and is sensitive to multiparton interactions, colour connections and hadronisation. Charged-particle multiplicities, transverse-momentum spectra and azimuthal distributions are useful observables for characterising this activity.

Bottomonium production provides a complementary probe of these correlations. The three lowest bottomonium states, \UpsOne, \UpsTwo and \UpsThree, originate from the same bottom-quark pair but have different masses and binding energies. Comparing the charged-particle activity accompanying the different states therefore provides information on the connection between bottomonium formation and the surrounding hadronic environment.

Previous measurements by the CMS Collaboration observed a decrease in the relative production of excited \Upsn states with increasing charged-particle multiplicity in \pps collisions \cite{CMS:eventactivity1,CMS:eventactivity2}. The ATLAS measurement presented here approaches this question by using the \Upsn state as a tag and measuring the charged particles accompanying it. Their multiplicity, transverse momentum and azimuthal separation from the
\Upsn direction are studied as functions of \ptmumu. The results use the complete Run~2 \pps data set at $\s=13~\TeV$, corresponding to 139~\ifb~\cite{ATLASCONF}.

\section{Data set, event selection and corrections}
The analysis uses the complete Run~2 \pps data set collected by ATLAS during 2015--2018 at $\s=13~\TeV$, corresponding to an integrated luminosity of 139~\ifb. Periods with 50 or more average simultaneous interactions are excluded, resulting in an average pile-up of approximately 33.7~\cite{ATLAS:DQ}.

The \Upsn candidates are reconstructed through $\Upsn\rightarrow\mu^+\mu^-$ using several dimuon triggers with thresholds between 4 and 11~\GeV. Muons are required to satisfy $\pT^\mu>4~\GeV$ and $|\eta^\mu|<2.4$, with higher thresholds applied where required by the trigger configuration~\cite{ATLAS:muontrig,ATLAS:muonreco}. Oppositely charged muon pairs are selected in the invariant-mass range $8.2<m^{\mu\mu}<11.8~\GeV$ and are required to satisfy $|y^{\mu\mu}|<1.6$.

Primary charged particles are selected with $0.5<\ptch<10~\GeV$ and $|\eta|<2.5$, using standard ATLAS track-quality requirements~\cite{ATLAS:tracking}. Only tracks within 0.5~mm of the averaged vertex position in the transverse direction and within 0.75~mm of the vertex associated with the muons in the longitudinal direction are considered.  The two muon tracks forming the \Upsn candidate are excluded from the charged-particle observables.

Detector acceptance, reconstruction and trigger efficiencies are corrected using simulation and data-driven scale factors. Pile-up contributions are estimated and subtracted using the procedure developed for the ATLAS $Z$-tagged UE measurement~\cite{ATLAS:Ztag}. The detector response is simulated using \Pythia with the A14 tune and CTEQ6L1 parton distribution functions, followed by a \Geant detector simulation~\cite{Pythia8,A14,Geant4}. An alternative colour-reconnection model is also considered as a cross-check~\cite{CR}.

\section{Extraction of the \Upsn-tagged distributions}
The three \Upsn peaks overlap and are accompanied by a smooth dimuon background, so no single mass window isolates an individual state. The charged-particle distributions are therefore measured in five dimuon mass intervals spanning the three signal regions and the two background-dominated side-bands.

The dimuon mass spectrum is fitted separately in each \ptmumu interval and trigger category using three signal components for \UpsOne, \UpsTwo and \UpsThree, together with a smooth third-degree polynomial background. Each signal is described by a Crystal Ball and Gaussian component; the low-mass tail accounts for final-state radiation. The signal yields and background normalisation are determined from the mass fit.

Figure~\ref{fig:massmult} illustrates the extraction procedure. The left panel shows the dimuon invariant-mass distribution for $10\le \ptmumu<12~\GeV$, together with the fit used to determine the \Upsn yields. The right panel shows the charged-particle multiplicity distributions in the mass intervals used for the signal and background separation. The pile-up contribution has a similar shape across the intervals and is controlled in the
subtraction procedure~\cite{ATLASCONF}.

\begin{figure}[tb!]
\centering
\includegraphics[width=0.8\textwidth]{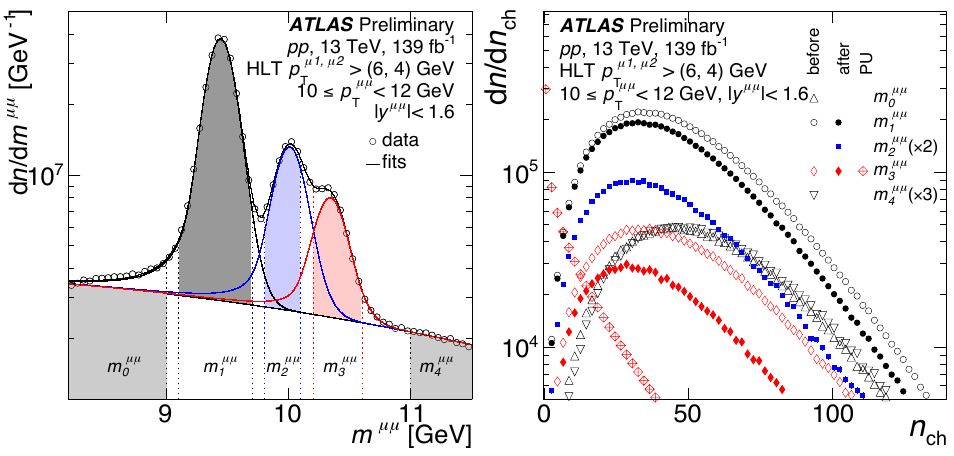}
\caption{Invariant-mass distribution and fit used to extract the \Upsn yields (left), and charged-particle multiplicity distributions in the mass intervals used for signal and  background separation (right), for  $10 \le \ptmumu<12~\GeV$.}
\label{fig:massmult}%
\end{figure}
The background charged-particle distribution is determined from the two side-band regions and interpolated beneath the three signal regions. The resulting distributions are combined using the signal fractions and cross-contamination obtained from the mass fit to determine the charged-particle observables for each \Upsn state. Pseudo-experiments show deviations from the input distributions of at most 1$\%$.

The dominant systematic uncertainties arise from charged-track reconstruction, \Upsn signal extraction and pile-up subtraction. Tracking uncertainties include detector material and reconstruction effects~\cite{ATLAS:trackingsyst}. Additional uncertainties are obtained by varying the signal and background models, studying detector stability and evaluating possible contributions from mis-reconstructed tracks. The total uncertainty on the mean charged-particle multiplicity is 0.5-1.3 charged particles, while the uncertainty on differences between states is 0.06-0.9 charged particles, with substantial cancellation of correlated uncertainties.

\section{Results}
\subsection{Transverse momentum and azimuthal distributions}
Figure~\ref{fig:ptdphi} shows the charged-particle transverse-momentum and azimuthal distributions for different \ptmumu intervals. For \UpsOne, the charged-particle spectrum becomes harder with increasing \ptmumu, while the azimuthal distribution develops near-side and away-side peaks characteristic of correlated jet activity.
\begin{figure}[tb!]
\centering
\includegraphics[width=0.9\textwidth]{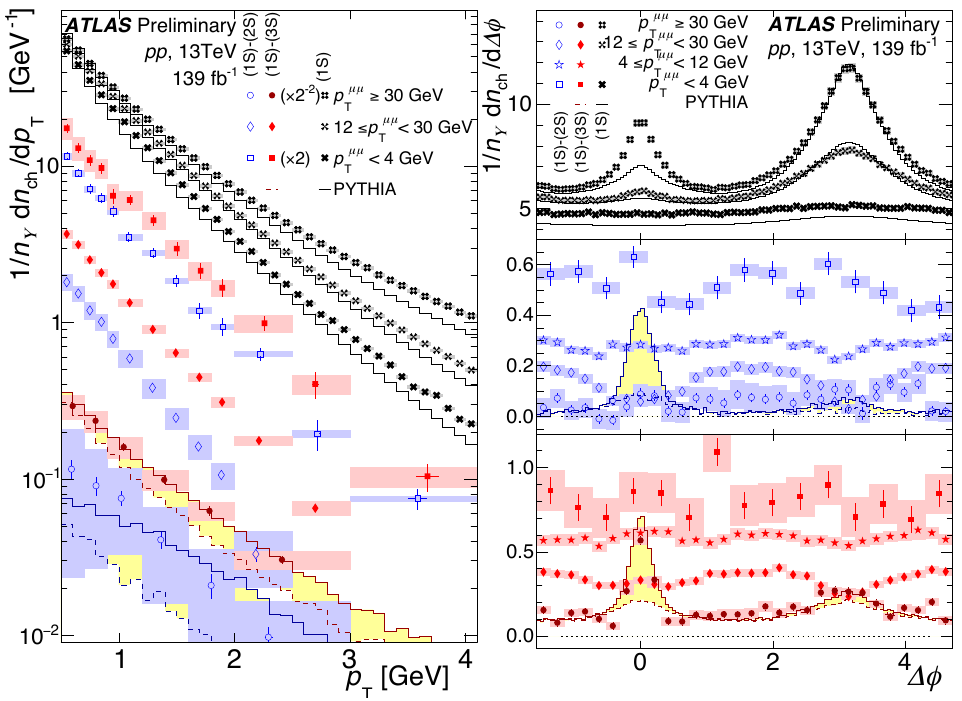}
\caption{Charged-particle transverse-momentum (left) and azimuthal separation from the \Upsn direction (right) for different \ptmumu intervals. The \UpsOne distributions and the
  \UpsOne-\UpsTwo and \UpsOne-\UpsThree differences are compared with \Pythia predictions.}
\label{fig:ptdphi}
\end{figure}
\Pythia qualitatively reproduces the jet-like structures, but underestimates the low-\ptch activity and the near-side contribution. The lowest \ptmumu interval, below 4~\GeV, is most sensitive to soft event activity.

The subtracted distributions remain positive over the measured \ptch range. Their transverse-momentum dependence is more similar to the softer \UpsOne distribution at low \ptmumu than to the corresponding higher-momentum distribution. The subtracted \dphi distributions are approximately flat and remain positive around $\dphi=\pm\pi/2$, where the contribution from correlated jets is reduced. This indicates that the observed differences between the states contains a substantial soft component.

At higher \ptmumu, additional azimuthal non-uniformities appear. These may receive contributions from feed-down from higher bottomonium states, including $\chi_b$ decays, as well as differences in jet activity.

\subsection{Mean charged-particle multiplicity}
The mean charged-particle multiplicity is measured for charged particles with $0.5 < \ptch < 10~\GeV$ and $|\eta|<2.5$. Figure~\ref{fig:meannch} shows the multiplicity as a function of \ptmumu for the three bottomonium states. 
\begin{figure}[tb!]
\centering
\includegraphics[width=0.5\textwidth]{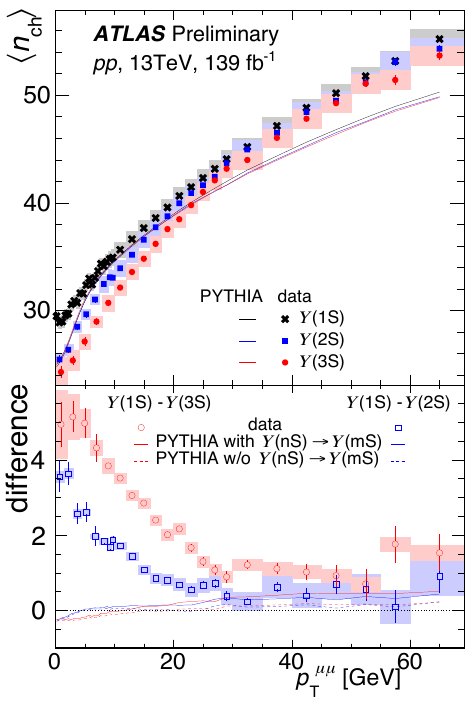}
\caption{Mean number of charged particles as a function of \ptmumu for events containing \UpsOne, \UpsTwo and \UpsThree mesons. The lower panel shows the differences between the excited and ground states. \Pythia predictions with and without feed-down contributions are shown for comparison.}
\label{fig:meannch}%
\end{figure}
A clear hierarchy is observed among the three bottomonium states. In the lowest measured \ptmumu interval, events containing \UpsOne measons have a mean charged-particle multiplicity of $29.5 \pm 0.7$. The corresponding multiplicities for \UpsTwo and \UpsThree are lower by $3.6 \pm 0.4$ and $4.9 \pm 1.1$, respectively. These differences corresponds to relative reductions of $12 \pm 1\%$ and $17 \pm 4\%$ with respect to the \UpsOne sample.

The \ptmumu dependences of the two differences are similar within the experimental uncertainties. The difference decreases as \ptmumu increases, but remains above zero even in the highest measured momentum interval. The uncertainty is larger for \UpsThree, particularly at low \ptmumu, beacuse of its smaller production yield.

In contrast to the data, \Pythia predicts very similar charged-particle multiplicities for the three \Upsn states. Including or excluding \Upsn feed-down contributions does not reproduce the observed hierarchy. The conclusion is also unchanged when the colour-reconnection scheme is varied. The small positive differences predicted at higher \ptmumu arise mainly from the different bottomonium masses and the feed-down decays.

The measured multiplicity hierarchy therefore provides a strong constraint on the modelling of bottomonium production and its correlation with the surrounding soft event activity. In particular, the data indicate that events producing the excited bottomonium states are accompanied by systematically lower charged-particle activity than events producing the ground states, with the largest difference observed in the lowest \ptmumu region.

\section{Summary}
The ATLAS Collaboration has measured charged-particle multiplicity, transverse-momentum and azimuthal distributions in \pps collisions at $\s=13~\TeV$ containing \UpsOne, \UpsTwo and \UpsThree mesons. The measurement uses the complete Run 2 data set of 139\ifb, with the bottomonium states reconstructed through their dimuon decays.

A clear hierarchy in charged-particle activity is observed. In the lowest measured \ptmumu interval, the \UpsOne sample has a mean charged-particle multiplicity of $29.5\pm0.7$, while the \UpsTwo and \UpsThree samples contain $3.6\pm0.4$ and $4.9\pm1.1$ fewer charged particles. These corresponds to relative differences of $12\pm1\%$ and $17\pm4\%$, respectively. The difference decrease with increasing \ptmumu, but remain positive over the measured range.

The charged-particle kinematic distributions show that the difference is not solely associated with the jet activity balancing the \Upsn momentum. In particular, the subtracted distributions retain a sizeable soft component, including in the azimuthal transverse region. The observed multiplicity hierarchy is not reproduced by the considered \Pythia models, and changing the colour-reconnection prescription does not remove the discrepancy. 

The results provide constraints on models of bottomonium production, hadronisation and the interplay between quarkonium formation and soft QCD activity. The observed state dependence of charged-particle activity, particularly at low \ptmumu, motivates further studies of the relationship between bottomonium production and the hadronic envirnment in high-multiplicity \pps collisions.


\end{document}